\documentclass[aps,pre,showpacs,twocolumn]{revtex4}

\usepackage{amsmath,epsfig,amsfonts,graphicx,
amsgen,amsbsy,amsopn,amstext,amssymb}

\begin{document}

\title{Modulational Instability in Nonlinearity-Managed Optical Media}

\author{Martin Centurion$^{1}$, Mason A. Porter$^{2}$, Ye Pu$^3$,
P. G. Kevrekidis$^4$, D. J. Frantzeskakis$^5$ and Demetri Psaltis$^{3}$}
\affiliation{
$^{1}$Max Planck Institute for Quantum Optics, 85748 Garching, Germany \\
$^2$Department of Physics and Center for the Physics of Information, California Institute of Technology, Pasadena, CA  91125, USA
\\
$^3$Department of Electrical Engineering, California Institute of Technology, Pasadena, CA  91125, USA
\\
$^4$Department of Mathematics and Statistics, University of Massachusetts, Amherst MA 01003-4515, USA
\\
$^5$Department of Physics, University of Athens, Panepistimiopolis, Zografos, Athens 15784, Greece }

\begin{abstract}
We investigate analytically, numerically, and experimentally the modulational instability in a layered, cubically-nonlinear (Kerr) optical medium that consists of alternating layers of glass and air. We model this setting using a nonlinear Schr\"odinger (NLS) equation with a piecewise constant nonlinearity coefficient and conduct a theoretical analysis of its linear stability, obtaining a Kronig-Penney equation whose forbidden bands correspond to the modulationally unstable regimes. We find very good {\it quantitative} agreement between the theoretical analysis of the Kronig-Penney
equation, numerical simulations of the NLS
equation, and the
experimental results for the modulational instability. Because of the
periodicity in the evolution variable arising from the layered medium, we find multiple instability regions rather than just the one that would occur in uniform media.
\end{abstract}



\maketitle

\section{Introduction}

The modulational instability (MI) is a destabilization mechanism for plane waves that results from the interplay between nonlinear and dispersive effects \cite{cross,scott}. It leads to delocalization in momentum space and, in turn, to localization in position space and the formation of localized (solitary-wave) structures. The MI arises in many physical settings, including fluid dynamics (where it is also called the ``Benjamin-Feir instability'') \cite{benjamin67}, nonlinear optics \cite{ostrovskii69,hasegawabook,Agrawal}, plasma physics \cite{taniuti68,hasegawa}, and other areas. Recently, it has also taken center stage in atomic physics in studies of atomic Bose-Einstein condensates (BECs) \cite{wu01,kon,ourpra,salas2,brand04} (see also \cite{pk1} for a review).  In all of the above settings, the MI is one of the principal mechanisms leading to the emergence of localized ``coherent nonlinear structures'' and the formation of bright solitary waves (and trains of solitary waves).

The MI was originally analyzed in uniform media, predominantly in the framework of
the nonlinear Schr{\"o}dinger equation (NLS) \cite{sulem}, where a focusing nonlinearity leads to MI for sufficiently large plane-wave amplitudes (for a given wavenumber) or sufficiently small wavenumbers (for a given amplitude) \cite{hasegawabook}. More recently, several very interesting experimentally relevant settings with (temporally and/or spatially) {\it nonuniform} media have emerged. Research in this direction includes the experimental observation of bright matter-wave soliton trains in BECs \cite{streck02}, which were induced by a temporal change of the interatomic interaction from repulsive to attractive through Feshbach resonances. Subsequent theoretical work demonstrated how this effective change of the nonlinearity from defocusing to focusing leads to the onset of MI and the formation of soliton trains \cite{ourpra,salas2,brand04}. Such soliton trains can also be induced in optical settings, as has been demonstrated, for example, in the context of birefringent dispersive media \cite{wabnitz88,wabnitz99}.

The theme of the present article is the analysis of the MI in a setting with periodic nonuniformities. In the context of nonlinear science, such studies date back to the Fermi-Pasta-Ulam problem \cite{scott,focus} and have continued ubiquitously in numerous disciplines. Periodic nonuniformities arise not only in the physical sciences (optics, atomic physics, solid-state physics, etc.) but also in biology in, for example, DNA double-strand dynamics \cite{peyrard}. In optics, prominent areas in which periodicity takes center stage include the study of photonic crystals \cite{solj} and optically-induced lattices in photorefractive crystals \cite{efrem}. Additionally, there have been recent experimental observations of the MI in spatially periodic optical media (waveguide arrays) \cite{dnc0}. In solid-state physics, periodicity has been prevalent in the study of superconducting Josephson junctions \cite{ustinov}, among other topics. Similar studies have been reported in atomic physics in, for example, BECs confined in spatially periodic optical potentials (so-called ``optical lattices'') \cite{wu01,kon,smerzi,cata}. A key feature in such settings is that MI can even occur for defocusing nonlinearities for certain wavenumber bands, as was originally suggested in Ref.~\cite{kivsharpeyrard}.

While the aforementioned results pertain to {\it spatially} inhomogeneous settings, in which the periodicity lies in the transverse dimensions, we discuss in this article the theoretical analysis and the first experimental realization of MI in a setting that is periodic in the {\it evolution variable}. In the optical setting discussed here, this variable describes the propagation distance (it represents time in the framework of BECs). Such settings were initially proposed in the context of optical fiber communications, through so-called ``dispersion management'' techniques (which induce periodic changes in the group-velocity dispersion) \cite{progress,turit,nd,kutzmod,kumar}), and have since also been studied for ``nonlinearity management'' (in which the Kerr nonlinearity is periodic in the propagation variable) \cite{luc1,isaac,wise,centurion}. They have also been investigated in BECs through so-called ``Feshbach Resonance Management" (FRM) schemes \cite{FRM,abdul,ueda,monte,vpghill,abdul3,abdul96,zoi2} (see also the review \cite{borisbook}).

In the present paper, which provides a detailed description of work we reported in a recent Letter \cite{mishort}, we discuss an experimental investigation of MI in a layered optical medium and its quantitative comparison with analytical and numerical results. The medium consists of alternating layers of glass and air, through which the optical waves traverse. We model the dynamics of our experiment using an NLS equation with piecewise constant nonlinearity coefficients and a dissipation mechanism that accounts for reflective losses at the glass--air interfaces. We obtain very good {\it quantitative} agreement between our theoretical analysis, numerical simulations, and the experimental results. The tractability of the theoretical analysis in our setting stems from the piecewise-constant nature of the material coefficients. In particular, this leads to an ordinary differential equation (ODE) in the linear stability analysis of the plane waves that is reminiscent of the Kronig-Penney (KP) model of solid-state physics \cite{kittel}. This model is a special case of a Hill equation \cite{magnus}, whose coefficients are periodic in the evolution variable, leading to a band structure for its eigenvalues. In the present context, the forbidden bands of the KP model correspond to the instability regimes. This observation allows us to compute the bands of modulationally unstable wavenumbers of perturbation (on top of the plane-wave solution) semi-analytically and consequently to compare our experimental findings not only with the numerical simulations of the NLS equation but also with the theoretical analysis of the KP model.  Similar analyses have been performed, from a theoretical point of view, for dispersion-managed optical fibers
\cite{nd,kutzmod,kumar}.

The remainder of this paper is organized as follows.
First, we present our theoretical model governing pulse propagation in layered Kerr media and we analyze it mathematically.  We then discuss our numerical procedures and experimental setup, present our main results, and provide additional discussion on several technical points. Finally, we present our conclusions and some future challenges.



\section{Theoretical analysis}

\subsection{Analytical results}
Our theoretical model for the propagation of the optical beam in the layered nonlinear medium incorporates the dominant dispersive and Kerr effects for each of the two media. Accordingly, by employing the slowly-varying envelope approximation to the Maxwell equations \cite{photon}, we derive the following
NLS equation:
\begin{align}
    i\frac{\partial u}{\partial \zeta} &= -\frac{1}{2}{\nabla}^2_\perp u - |u|^2u \,, \quad 0 < \zeta < \tilde{l}\, \quad \mbox{(glass)}\,, \notag \\
        i\frac{\partial u}{\partial \zeta} &= -\frac{1}{2}\frac{n_0^{(1)}}{n_0^{(2)}}{\nabla}^2_\perp u - \frac{n_2^{(2)}}{n_2^{(1)}}|u|^2u \,, \quad \tilde{l} < \zeta < \tilde{L}\, \quad \mbox{(air)}\,. \label{nls}
\end{align}
In Eq.~(\ref{nls}), space is rescaled by the wavenumber $k^{(1)} = 2 \pi n_0^{(1)}/ \lambda$ (where $\lambda$ is the source wavelength); that is, $(\xi,\eta,\zeta)=k^{(1)}\! \times \!(x,y,z)$.
The complex field $u$ is given by $u=(n_2^{(1)}/n_0^{(1)})^{1/2} E$, where $E$ is the electric field envelope.
The superscript $(j)$ denotes the medium, with $j=1$ for glass and $j=2$ for air. As is well-known, the linear parts $n_0^{(j)}$ of the refractive indices of glass and air take the values $n_0^{(1)}=1.5$ and $n_0^{(2)}=1$, respectively.  The nonlinear parts (i.e., the Kerr coefficients), $n_2^{(j)}$, are $n_2^{(1)}=3.2\times10^{-16}$ cm$^2$/W and $n_2^{(2)}=3.2\times10^{-19}$ cm$^2$/W \cite{Boyd}.

One can also incorporate the transmission losses at each slide into the system (\ref{nls}).  It is then written
%
\begin{equation}
    i \frac{\partial u}{\partial \zeta}=-\frac{1}{2} D(\zeta) \nabla^2 u - N(\zeta) |u|^2 u - i \gamma(\zeta) u\,,
\label{nls1}
\end{equation}
where $D(\zeta) = 1$, $N(\zeta) = 1$ in glass and $D(\zeta) = n_0^{(1)}/n_0^{(2)}$, $N(\zeta) = n_2^{(2)}/n_2^{(1)}$ in air.
The last term in Eq.~(\ref{nls1})
describes the transmission losses at each slide.
The loss rate $\gamma(\zeta)$ is given by
\begin{equation}
    \gamma(\zeta) = \alpha \sum_{n = 1}^M \delta(\zeta - \zeta_n)\,,
\end{equation}
where $M$ is the number of glass--air interfaces at which losses occur and $\zeta_n$ is the location of the $n$-th interface. The prefactor $\alpha$ is determined by the constraint that the power $P$ after an interface is a factor $r$ (which for our experiments is typically $0.99$) times the power before the interface.
Because $dP/d\zeta = -2\gamma P$, the parameter $\alpha$ satisfies the equation $\exp(-2\alpha) = r$.

To examine the onset of MI, we consider plane-wave solutions of Eq. (\ref{nls}), which are uniform in the tranverse spatial variables ($\xi$ and $\eta$). Using the transformation
\begin{equation}
    v = u\exp\left[\int^\zeta \gamma(\zeta')d\zeta'\right]\,,
\end{equation}
we obtain the equation
\begin{equation}
    i\frac{dv}{d\zeta} = -N(\zeta) \exp\left[-2 \int^\zeta \gamma(\zeta')d\zeta'\right]|v|^2v \,,
\end{equation}
where we note that the Laplacian term in Eq.~(\ref{nls}) is identically zero. Transforming to polar coordinates, $v = Re^{i\theta}$, we subsequently obtain
\begin{align}
    i\frac{dR}{d\zeta} - \frac{d\theta}{d \zeta}R &= -N(\zeta)\exp\left[-2\alpha\sum_{n=1}^MH(\zeta - \zeta_n)\right]R^3\,, \notag \\
    \frac{dR}{d\zeta} &= 0\,,  \label{polar}
\end{align}
where $H(\zeta')$ is the Heaviside step function.  Equation (\ref{polar}) implies $R = R(0) \equiv A_0$ and
\begin{equation}
    \frac{d\theta}{d\zeta} = A_0^2N(\zeta)\exp\left[-2\alpha\sum_{n=1}^MH(\zeta - \zeta_n)\right]\,.
\end{equation}
This yields an analytical expression for the plane-wave solutions of Eq.~(\ref{nls1}),
\begin{equation}
   u_0(\zeta)=A_0 e^{-\int^{\zeta} \gamma(\zeta') d \zeta'} e^{i A_0^2 \int^{\zeta} N(\zeta')
    \Gamma(\zeta') d\zeta'}\,, \label{nls2}
\end{equation}
where 
	$\Gamma(\zeta') \equiv  \exp({-2\int^{\zeta'} \gamma(\tilde{\zeta}) d\tilde{\zeta}})$.

To perform a linear stability analysis of these plane waves, we consider a spatial perturbation of (\ref{nls2}) given by
\begin{equation}
    u = u_0(\zeta) \left[1+ w(\zeta) \cos(k_{\xi} \xi) \cos(k_{\eta} \eta)\right]\,, \label{linear}
\end{equation}
where $w$ is a small (Fourier-mode) perturbation with wavevector $(k_\xi,k_\eta)$. We use the notation $w = \varepsilon \tilde{w}$ and insert the expression (\ref{linear}) into Eq.~(\ref{nls1}) to obtain
\begin{align}
    &i\frac{du_0}{d\zeta}(\zeta)[1 + \varepsilon\tilde{w}(\zeta)
\cos(k_{\xi} \xi) \cos(k_{\eta} \eta)] \notag \\ &\quad + i\varepsilon u_0(\zeta)\frac{d\tilde{w}}{d\zeta}(\zeta)\cos(k_\xi\xi)\cos(k_\eta\eta)  + O(\varepsilon^2) \notag \\
    &= -\frac{1}{2}\varepsilon u_0(\zeta)D(\zeta)(-k_\xi^2 - k_\eta^2)\tilde{w}(\zeta)\cos(k_\xi\xi)\cos(k_\eta\eta) \notag \\ &\quad - i\gamma(\zeta)u_0(\zeta)[1+ \tilde{w}(\zeta) \cos(k_{\xi} \xi) \cos(k_{\eta} \eta)] \notag \\ &- N(\zeta)|u_0(\zeta)|^2u_0(\zeta)\left[1 + 2\varepsilon \tilde{w}(\zeta) \cos(k_{\xi} \xi) \cos(k_{\eta} \eta)\right] \times \notag \\ &\quad \times \left[1 + \varepsilon \tilde{w}^*(\zeta) \cos(k_{\xi} \xi) \cos(k_{\eta} \eta)\right]  + O(\varepsilon^2) \,,
    \label{expand}
\end{align}
where $\tilde{w}^*$ denotes the complex conjugate of $\tilde{w}$. We then equate the terms in (\ref{expand}) order by order in powers of $\varepsilon$. By construction, the $O(1)$ terms cancel out. The $O(\varepsilon)$ terms give
\begin{align}
    &i\frac{du_0}{d\zeta}(\zeta)\tilde{w}(\zeta) + iu_0(\zeta)\frac{d\tilde{w}}{d\zeta}(\zeta) = \frac{1}{2}(k_\xi^2 + k_\eta^2)D(\zeta)u_0(\zeta)\tilde{w}(\zeta) \notag \\ &\quad - i\gamma(\zeta)u_0(\zeta)\tilde{w}(\zeta) - N(\zeta)|u_0(\zeta)|^2u_0(\zeta)(2\tilde{w} + \tilde{w}^*)\,. \label{inter}
\end{align}
Dividing both sides of (\ref{inter}) by $u_0(\zeta)$ and using the equation
\begin{equation*}
    i\frac{1}{u_0(\zeta)}\frac{du_0}{d\zeta}(\zeta) = -N(\zeta)|u_0(\zeta)|^2 - i\gamma(\zeta),
\end{equation*}
yields
\begin{align}
    &i\frac{d\tilde{w}}{d\zeta}(\zeta) = \frac{1}{2}(k_\xi^2 + k_\eta^2)D(\zeta)\tilde{w}(\zeta) \notag \\ &\quad - N(\zeta)|u_0(\zeta)|^2\tilde{w}(\zeta) - N(\zeta)|u_0(\zeta)|^2\tilde{w}^*(\zeta)\,.
\end{align}

Decomposing $\tilde{w}$ (and hence $w$) into real and imaginary parts, $\tilde{w}=F+ i B$, we obtain a linear system of ODEs,
\begin{align}
    \frac{dF}{d\zeta} &= \frac{1}{2}\bar{k}^2D(\zeta)B\,,  \\
    \frac{dB}{d\zeta} &= -\frac{1}{2}\left[\bar{k}^2D(\zeta) +  2N(\zeta) |u_0(\zeta)|^2\right]F\,, \notag
\end{align}
where $\bar k^2=k_{\xi}^2+k_{\eta}^2$.  We rewrite the above system as a single second-order equation,
\begin{eqnarray}
    \frac{d^2 F}{d \zeta^2}&=&\frac{1}{D(\zeta)}\frac{d D}{d \zeta}(\zeta)\frac{d F}{d \zeta} \nonumber \\
&+& \left[-\frac{1}{4} \bar k^4 D(\zeta)^2 + N(\zeta) \bar k^2 D(\zeta) |u_0(\zeta)|^2\right] F. \label{nls3}
\end{eqnarray}
The transformation $F = g D^{1/2}$ then yields a Hill equation,
\begin{align}
    \frac{d^2g}{d\zeta^2} &= \left[S(\zeta) + \frac{3}{4}\frac{1}{D(\zeta)^2}\left(\frac{dD}{d\zeta}(\zeta)\right)^2 - \frac{1}{2}\frac{1}{D(\zeta}\frac{d^2D}{d\zeta^2}(\zeta)\right]g\,, \notag \\
    S(\zeta) &= \frac{1}{2}\bar k^2D(\zeta)\left[2N(\zeta)|u_0(\zeta)|^2 - \frac{1}{2}\bar k^2D(\zeta)\right]\,. \label{nls3a}
\end{align}

Hereafter, we make some simplifying assumptions that we will justify based both on the experimental system parameters and on a detailed comparison of our results with the numerical simulations and the experimental findings. In particular, in investigating Eq.~(\ref{nls3a}), we ignore the losses at the interfaces so that its solution is periodic and Bloch's theorem can be employed. Given that the transmission for each slide is approximately $99 \%$, this is a very good approximation. Additionally, while it is possible to analyze Eq.~(\ref{nls3}) directly (if losses are ignored), it is sufficient in modeling our experiments to exploit the weak variation of $D(\zeta)$ and substitute $D(\zeta)$ with its average. In this case, the transformation to Eq.~(\ref{nls3a}) is no longer necessary, and we proceed using Eq.~(\ref{nls3}). As will be discussed below, this assumption has hardly any effect on the results from Eq.~(\ref{nls3}). Under this additional simplification, Eq.~(\ref{nls3}) is a Hill equation \cite{magnus} equivalent to the well-known Kronig-Penney model from solid-state physics (originally aimed as a prototypical description of an electron moving through a crystal lattice in a solid) \cite{kittel} for the piecewise-constant nonlinearity coefficient under consideration. As discussed in Ref.~\cite{zoi2}, such a periodic potential allows one to use Bloch's theorem to obtain an analytical solution in both glass and air.  Consequently, one may write \cite{kittel,675}
\begin{equation}
    F(\zeta +\tilde{L}) = \exp\left(-i\omega\tilde{L}\right)F(\zeta)\,,
\end{equation}
where $\omega$ is the Floquet multipler.
The analytical solution for $F(\zeta)$ is
\begin{align}
    F &= a_1e^{i s_1\zeta} + b_1 e^{-i s_1\zeta}        \,,
        \quad 0 < \zeta < \tilde{l}\, \quad \mbox{(glass)}\,, \notag \\
    F &= a_2e^{i s_2\zeta} + b_2e^{-i s_2\zeta} \,,
        \quad \tilde{l} < \zeta < \tilde{L}\, \quad \mbox{(air)}\,, \notag
\end{align}%
where $s_1^2=\bar k^2 D^{(1)} (\bar k^2 D^{(1)}/4-N^{(1)} |u_0|^2)$, $s_2^2=\bar k^2 D^{(2)} (\bar k^2 D^{(2)}/4-N^{(2)} |u_0|^2)$, and $\omega$ is the Floquet multiplier.  Because we are ignoring losses at the interfaces, $|u_0|^2 = |A_0|^2$ in the expressions for $s_1$ and $s_2$.

The continuity of $F$ and $dF/d\zeta$
at the glass--air boundaries (i.e., at $\zeta = \tilde{l}$ and $\zeta = \tilde{L}$) leads to matching conditions that are used to determine the constants of integration $a_1$, $a_2$, $b_1$, and $b_2$.
This
yields the following homogeneous $4 \times 4$ system of equations:
\begin{widetext}
\begin{equation}
    \begin{pmatrix}
       1 & 1 & -1 & -1 \\
       s_1& -s_1& -s_2 & s_2 \\
      e^{is_1\tilde{l}}   & e^{-is_1\tilde{l}}   & -e^{-i\omega\tilde{L}}e^{is_2(\tilde{l}-\tilde{L})} & -e^{-i\omega\tilde{L}}e^{-is_2(\tilde{l}-\tilde{L})} \\
     s_1e^{is_1\tilde{l}} & -s_1e^{-is_1\tilde{l}} &  -s_2e^{-i\omega\tilde{L}}e^{is_2(\tilde{l}-\tilde{L})}&   s_2e^{-i\omega\tilde{L}}e^{-is_2(\tilde{l}-\tilde{L})}\notag
    \end{pmatrix}
    \begin{pmatrix}
        a_1 \\ b_1 \\ a_2 \\ b_2
    \end{pmatrix}
    = 0\,,
\end{equation}
\end{widetext}
which possesses nontrivial solutions if and only if the determinant of the matrix vanishes. This gives the following solvability condition for $\omega$:
\begin{align}
        &\cos(\omega \tilde{L}) = -\frac{s_{1}^{2}+{s}_{2}^{2}}{2s_{1} {s}_{2}}\sin (s_{1}\tilde{l} )
    \sin [{s}_{2}(\tilde{L}-\tilde{l} )] \notag \\ &\quad +\cos (s_{1}\tilde{l} )\cos [{s}_{2}(\tilde{L}-\tilde{l} )] \equiv G(\bar k)\,,
\label{nls4}
\end{align}
similar to that obtained for plane-wave solutions of the NLS equation with piecewise constant dispersion management \cite{kutzmod}.
%
Because of the functional form of the left-hand-side of Eq.~(\ref{nls4}), real solutions for the Floquet exponent $\omega$ exist if and only if $|G(\bar k)| \leq 1$. Given the form of the solution for the perturbation $F(\zeta)$, this case corresponds to modulationally stable wavenumbers $\bar{k}$ that exhibit oscillatory behavior. On the other hand, for $|G(\bar k)|>1$, the solutions of Eq.~(\ref{nls4}) are imaginary, leading to an exponential growth in the real part of the perturbation $F(\zeta)$ which, in turn, indicates that such wavenumbers are modulationally unstable. Hence, the analogy to the original Kronig-Penney problem shows that the allowable energy zones are the ones corresponding to stable wavenumbers, whereas the forbidden energy zones
are the ones associated with MI.


\subsection{Numerical setup}
We perform direct numerical simulations of Eq.~(\ref{nls1}) using experimentally-determined parameters.  To confirm the validity of our simulations, we use two different algorithms: a beam-propagation, split-step code and a code using finite differences in the spatial variables and a fourth-order Runge-Kutta algorithm in the propagation direction.  We used the latter method to generate all of the figures shown in this paper and have verified that we obtain the same results using the former algorithm.

For the depicted figures, we used a grid with 1000 spatial nodes and scaled the spatial step size by the wavenumber for each simulation to ensure periodic boundary conditions in a domain encompassing 15 periods of the periodic initial condition [see, in particular, Eq.~(\ref{linear})].  We obtained the same results upon varying the number of grid points (up to 4000) and the number of periods. The typical grid spacing and evolution-variable step were approximately $0.3 \mu$m and $0.04 \mu$m, respectively.

In conducting our numerical experiments, we make two additional simplifications.  First, motivated by the experiment, we consider in our numerical simulations the one-dimensional dynamics along the direction of the modulation.  That is, we use $k_{\eta}=0$ and vary $k_{\xi}$.  Accordingly, we convert the experimental two-dimensional interference patterns recorded on the CCD camera (see the next section for the specifications) to one-dimensional ones by integrating along the direction orthogonal to the modulation.
 Second, we assume that the modulational dynamics of the (weakly decaying) central part of the Gaussian beam of the experiment is similar to that of a plane wave with the same intensity. We tested both of these assumptions and confirmed them both a priori through the dynamical evolution of our experimental and numerical results and a posteriori through their quantitative comparison.  Consequently, the initial wavefunctions in the numerical experiments take the form
\begin{equation}
    u = A_0 + \epsilon_0 e^{i k_\xi \xi} \,, \label{input}
\end{equation}
where $A_0 = \sqrt{I_0}$, and $I_0$ denotes the intensities used experimentally (see the discussion below).  We also use the experimental perturbation value of $\epsilon_0 = A_0/10$. We follow (and report) the subsequent evolution in both real and Fourier space.



\section{Experiments vs. Theory}

\subsection{Experimental setup}
\begin{figure}[tbp]
\centerline{\includegraphics[width=8 cm]{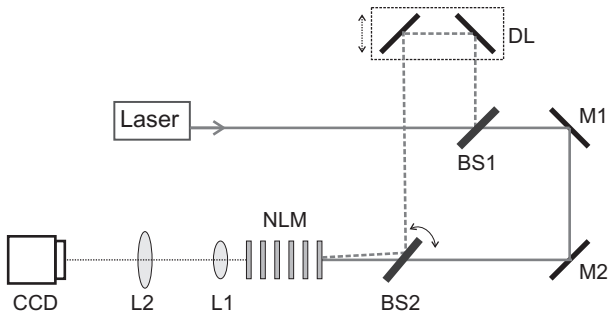} } \caption{Experimental setup. BS1 and BS2 are beam splitters, DL is a variable delay line, M1 and M2 are mirrors, NLM is the layered nonlinear medium, and L1 and L2 are lenses.} \label{setup}
\end{figure}

In our experiments, which are shown schematically in Fig.~\ref{setup}, we use an amplified Titanium:Sapphire laser to generate 150-femtosecond pulses with an energy of 2 mJ at a wavelength of $\lambda=800$ nm. The beam profile is approximately Gaussian with a full-width at half-maximum (FWHM) of 1.5 mm. The laser pulses are split into a pump and a reference using a beam splitter (BS1), with most of the energy in the pump pulse. After synchronization with a variable delay line (DL), the two pulses are recombined at a second beam splitter (BS2) and sent to the periodic nonlinear medium (NLM). This configuration allows us to control the relative angle in the propagation of the two beams.

The reference introduces a sinusoidal modulation in the intensity (i.e., an interference pattern), with the period determined by the relative angle between the two beams. We carefully tune the angle of the reference by rotating BS2 so that the two beams overlap while propagating through the NLM (at adjustable angles).  The NLM consists of six 1 mm thick quartz microscope slides separated by air gaps. The glass slides have an anti-reflection coating to minimize the loss (the reflection from each interface is 1\%). The loss due to back-reflections from the slides is included in our numerical simulations, and the effect of double reflections is negligible. In our experiments, we used structures with air gaps of 2.1 mm and 3.1 mm. We image the intensity pattern after the NLM (at the output face of the last quartz slide) on a CCD camera (Pulnix TM-7EX) using two lenses (L1 and L2) in a 4-F configuration with a magnification of $M = 8$. An image of the pump beam is used as a background and subtracted from the interference pattern to remove spatial nonuniformities that are not due to MI. The CCD camera captures the central region of the beam (0.6 mm $\times$ 0.8 mm). (Because the decay in this region of the Gaussian beam is weak, we approximate it as a plane wave in the theory and computations.)

We record the intensity pattern at the output of the NLM for three different values of the pump intensity: $I_{P0}=9.0\times10^{8}$ W/cm$^2$, $I_{P1}=9.0\times10^{10}$ W/cm$^2$ and $I_{P2}=1.3\times10^{11}$ W/cm$^2$.  In all three cases, the intensity of the reference beam is 1\% of that of the pump. We measure the effect of the nonlinearity by comparing the output for high ($I_{P1}$ and $I_{P2}$) versus low intensity ($I_{P0}$).  For low intensity, the propagation is essentially linear. In the nonlinear regime, if the spatial frequency of the modulation lies within an instability window, the amplitude of the reference wave will increase at the expense of the pump.


\subsection{Results}

The input field is given by Eq.~(\ref{input}),
where $A_0$ and $\epsilon_0$ are the amplitudes of the pump and reference beams, respectively, and $|\epsilon_0|^2 \ll |A_0|^2$. For linear propagation (low pump intensity, $I_{P0}$), the intensity pattern at the output of the NLM is approximately the same as that at the input.  That is, it is about
\begin{equation}
        I_{0}(\xi) = |A_0|^2 +|\epsilon_0|^2 + 2A_0 \epsilon_0 \cos(k_\xi\xi)\,.
\end{equation}
For the nonlinear case (high pump intensity, $I_{P1}$ and $I_{P2}$), the amplitude of the waves changes and higher spatial harmonics are generated. The intensity at the output of the NLM is approximately
%
\begin{equation}
    I_{NL}(\xi) = |A_1|^2 + 2 A_1 \epsilon_1 \cos(k_\xi\xi)+ 2A_1 \epsilon_2 \cos(2k_\xi\xi)+\cdots\,, \label{high}
\end{equation}
where $A_1$ and $\epsilon_{n}$ ($n=1, 2, \cdots$) are, respectively, the amplitudes of the pump beam and the $n$-th harmonic at the output of the NLM. The Fourier transform (FT) of Eq.~(\ref{high}) is
\begin{align}
    FT(I_{NL}) &= |A_1|^2 \delta (f_\xi)
\notag \\ &\quad + A_1 \epsilon_1 \left[ \delta\left(f_\xi - \frac{k_\xi}{2\pi}\right) +
\delta \left(f_\xi + \frac{k_\xi}{2\pi}\right) \right] \notag \\ &\quad + A_1 \epsilon_2 \left[ \delta \left(f_\xi - \frac{k_\xi}{\pi}\right) +
\delta\left(f_\xi + \frac{k_\xi}{\pi}\right) \right] + \cdots\,.
\end{align}
The FT peak height ratios (first-order:zeroth-order), $r_0 = \epsilon_0/A_0$ and $r_1=\epsilon_1/A_1$, are approximately equal to the amplitude ratios of the reference and pump waves.
(For the experimental value of $\epsilon_0 = A_0/10$, the error introduced by this approximation is roughly $1\%$.). In the linear case, the value of $r_0$ remains constant. However, in the nonlinear case, the value of $r_1$ depends on the length of the nonlinear medium.
We thus use the ratio $R = r_1/r_0$ as a diagnostic measure for both our experimental and numerical results, so that $R > 1$ indicates growth of the perturbation. This measurement is equivalent to the ratio $r_1(\zeta=\bar\zeta)/r_1(\zeta=0)$ (where $\bar\zeta$ is the scaled length of the nonlinear medium), which compares the amplitude of the reference wave at the output to that at the input, but it is more robust experimentally because it accounts for other linear effects (such as the limited coherence length and spatial overlap of the pulses) that can affect the strength of the peaks in the Fourier transform. Therefore, the value of $R$ reflects only the changes that are caused by the nonlinearity. In the numerical simulations, the peaks in the Fourier transform are sharp (one pixel), whereas they are broader in the experiments. Accordingly, when computing $R$ from the experimental data, we use the area under the peaks instead of the peak value.


\begin{figure}[tbp]
\centerline {\includegraphics[width=8.0cm]{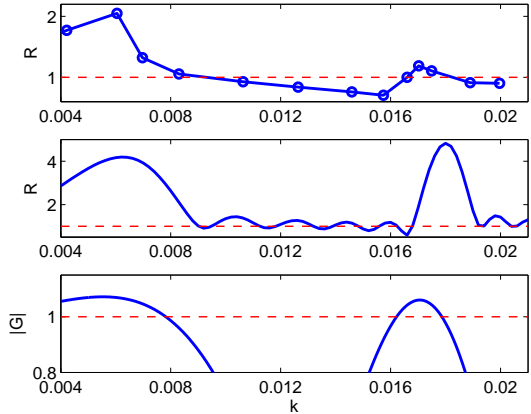}} \caption{(Color online) Comparison of experimental (top), numerical (middle), and analytical (bottom) results for the 1 mm glass--2.1 mm air configuration as a function of the dimensionless wavenumber $k$. For the diagnostics $R$ and $|G|$ (defined in the text), values larger than $1$ correspond to MI.} \label{mfig2}
\end{figure}

\begin{figure}[tbp]
\centerline{(a)\includegraphics[width=8.0cm]{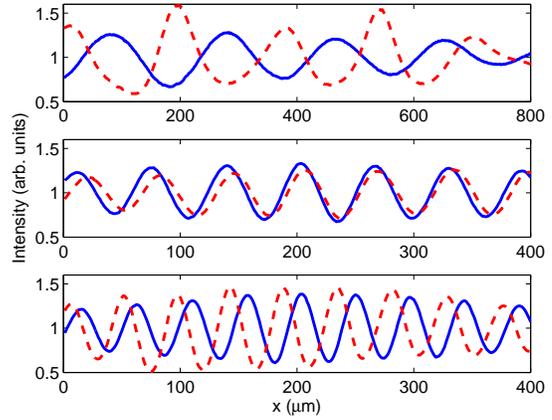}} \centerline{(b)\includegraphics[width=8.0cm]{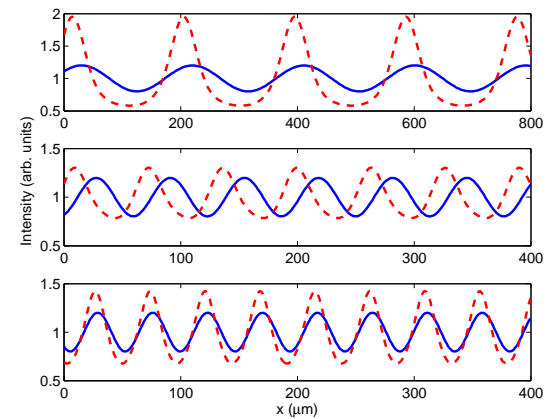}}
 \caption{(Color online) Comparison of experimental (a) and numerical (b) one-dimensional
intensity patterns at the output ($z=16.5$ mm) of the nonlinear medium with 1 mm glass--2.1 mm air for $k = 0.0042$ [top panels of (a) and (b)], $k = 0.0126$ (middle), and $k = 0.0170$ (bottom), corresponding to the first instability band, the following stable region, and the second instability band, respectively.  The dashed curves are for high intensity $(I_{P2})$ and the solid ones are for low intensity $(I_{P0})$. To facilitate comparisons between curves with different initial intensities, we scale the curves in this figure (to ``arbitrary units") using the condition that the mean of $|u(x,z)|^2$ is 1.} \label{mfig3}
\end{figure}

Figure \ref{mfig2} shows the ratio $R(k)$ for the structure with $1$ mm slabs of glass sandwiching $2.1$ mm air spacings, where $k=k_\xi \eta_0^{(1)}$ is the sine of the angle between the pump and reference beams.  There are two instability bands, quantified experimentally by $R > 1$, within the measurement range.  Similar to what has been shown computationally for dispersion-managed media \cite{nd,kumar}, the periodicity in the evolution variable from the layered (``nonlinearity-managed") medium induces a second instability band. Note that only a single band occurs in uniform media.
The maximum growth of the perturbation in the first and second bands appear at $k=6.0\times 10^{-3}$ and $k=1.70\times 10^{-2}$, with values of $R=2.05$ and $R=1.19$, respectively.  The increase in the modulation is clearly visible in the 1D intensity patterns (see Fig.~\ref{mfig3}). As indicated by the middle and bottom panels of Fig.~\ref{mfig2}, the positions of the instability bands are in very good agreement with both numerical and theoretical (indicated by the forbidden zones with $|G|>1$) predictions. The numerical simulations typically show a stronger instability than the experimental measurement; this results from the three-dimensional nature of the experiment that is not captured in the simulation.  In the experiment, the spatial and temporal overlap of the two beams decreases with increasing $k$, which leads to weakening of the higher-order peaks. Additionally, temporal dispersion leads to a reduction in the aggregate strength of the nonlinearity in the experiment. The small-amplitude ripples that appear in the numerical simulation within the stable region result from the finite number of periods in the propagation distance rather than from actual instabilities. We discuss this issue in further detail below.  Despite this difference, we stress that the simulations successfully achieve our primary goal of quantitatively capturing the locations of the instability windows.

Figure \ref{mfig3}(a) shows the normalized experimental one-dimensional intensity pattern at the output of the NLM for $I_{P2}$ (dashed curve) and $I_{P0}$ (solid curve). The three panels correspond to the cases $k=4.2\times 10^{-3}$ (top), which lies in the first instability band; $k=1.26\times 10^{-2}$ (middle), in the stable region separating the two forbidden zones; and $k=1.70\times 10^{-3}$ (bottom), which lies in the second instability band. The comparison of high and low intensity clearly shows the effect of the nonlinearity on the propagation. When the modulation of the input wave lies within the instability band, the amplitude of the modulation for the high intensity wave increases due to MI. We have observed an increase in the amplitude of the modulation both in the first and second instability bands, whereas the modulation remains practically unchanged in the stable region. We show the corresponding numerical intensity patterns in Fig.~\ref{mfig3}(b). As with the experimental results, the MI in the first and third panels causes the peaks in the intensity pattern to become higher and narrower, whereas this is not the case for the middle panel of the figure. We note in passing that a small increase in amplitude is discernible in the middle panels of Figs.~\ref{mfig3}(a,b). This is associated with the ripples mentioned above and will be discussed more extensively below.

\begin{figure}[tbp]
\centerline{\includegraphics[width=8.0cm]{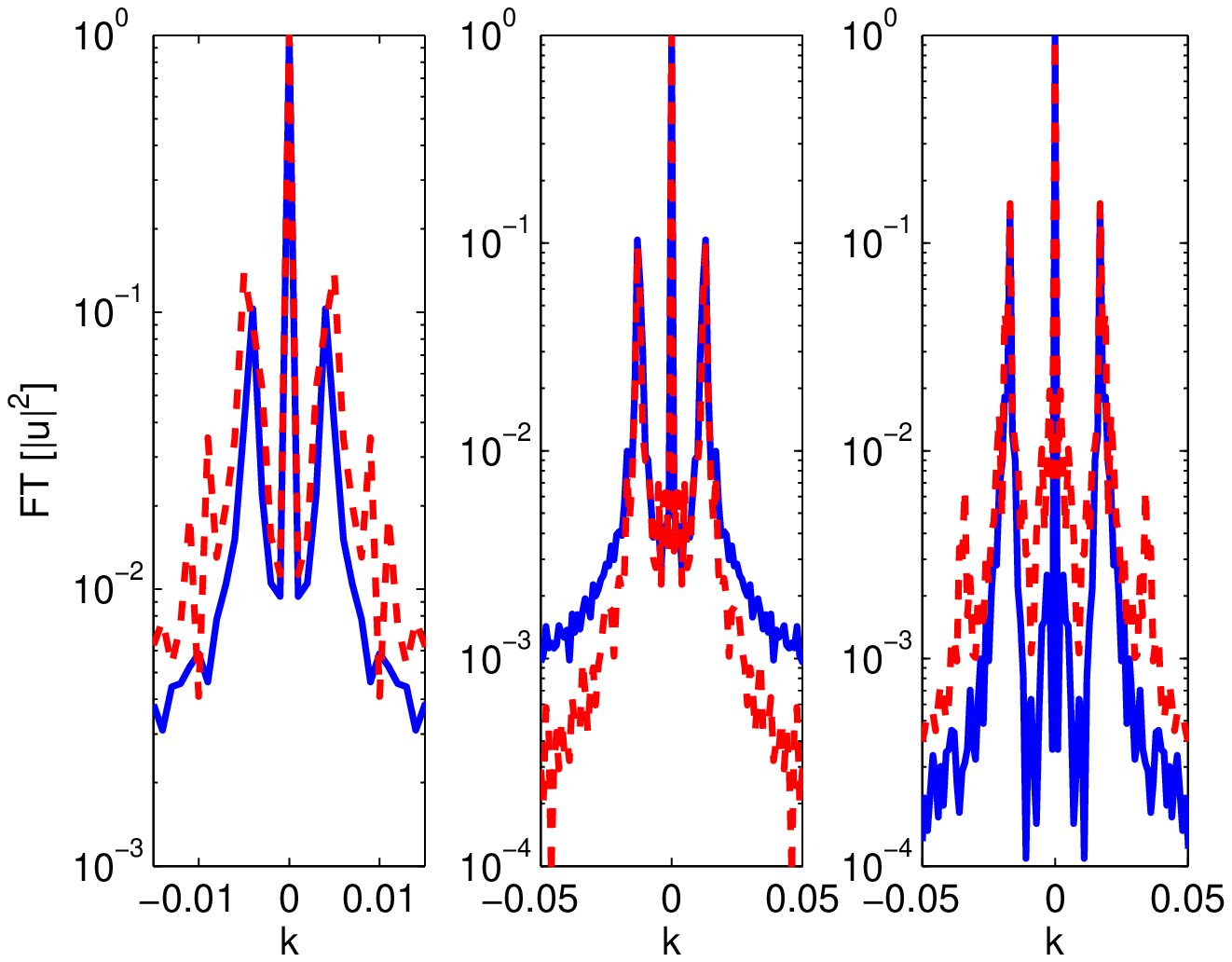}} \centerline{\includegraphics[width=8.0cm]{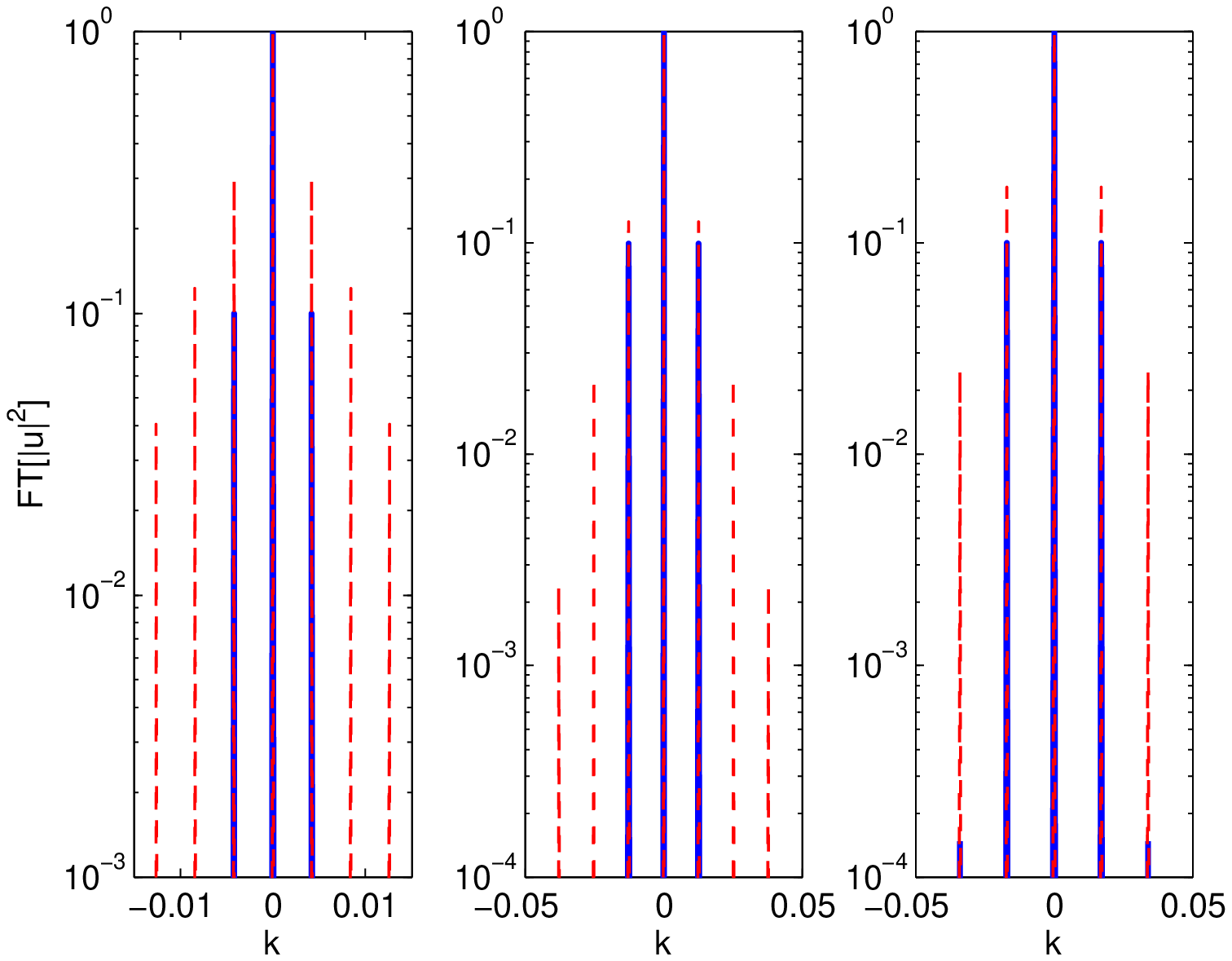}}
\caption{ (Color online) Experimental Fourier spectra (top panels) and numerical Fourier spectra (bottom panels) at the end of propagation ($z=16.5$ mm) of the layered structure with 6 glass slides (of thickness 1 mm), each pair of which sandwiches 2.1 mm of air. The left panels are for $k=0.0042$ (first instability band), the center ones are for $k=0.0126$ (stable region), and the right ones are for $k=0.0170$ (second instability band).  The dashed curves are for high intensity $(I_{P2})$ and the solid ones are for low  intensity $(I_{P0})$.} \label{mfig4}
\end{figure}

In Fig.~\ref{mfig4}, we show the Fourier transforms of the experimental (top panels) and numerical (bottom panels) intensity patterns at the output of our layered medium (i.e., at $z=16.5$ mm). Observe the appearance of higher spatial harmonics of the initial modulation in the regions of instability.
Such harmonics correspond to the narrowing of the peaks in the spatial interference pattern. The panels depict the wavenumbers $k = 0.0042$ (left), $k = 0.0126$ (center), and $k = 0.0170$ (right). The first-order peaks (the ones closest to $k=0$) correspond to the modulation of the input beam and are present for both low and high intensity. For high intensity, additional peaks appear at the higher harmonics for unstable wavenumbers of the modulation ($k = 0.0042$ and $k = 0.0170$). We also observed this harmonic generation in the numerical simulation (bottom panels), in good agreement with the experiments. In contrast, such harmonics are absent in the experimental results for $k$ within the modulationally stable regions ($R < 1$ in Fig.~\ref{mfig2}; see the top center panel of Fig.~\ref{mfig4}), again in agreement with the theoretical prediction. As with the ripples observed in Fig.~\ref{mfig2} and the weak intensity amplification of Fig.~\ref{mfig3}, the appearance of weaker harmonics in the numerical simulation for $k = 0.0126$ arises from the use of finitely many propagation periods in the numerical simulation; these result in a weak amplification even in modulationally stable cases. As we explain in detail below, incorporating additional propagation periods in the numerical evolution distinguishes with increasing clarity the modulationally stable and unstable regions.


\begin{figure}[tbp]
\centerline{\includegraphics[width=8.0cm]{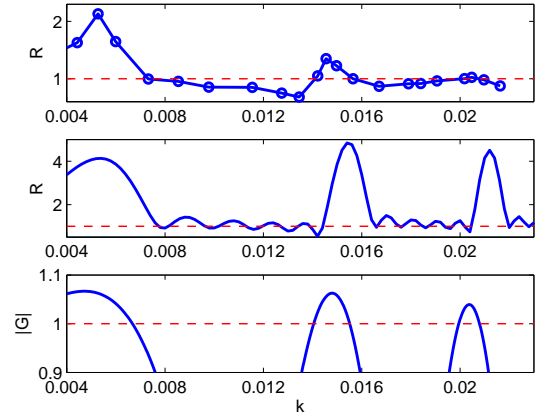} } \caption{(Color online) Same as Fig.~\ref{mfig2} but for the 1 mm glass--3.1 mm air configuration. Observe the presence of a third MI band.} \label{mfig6}
\end{figure}

Figure \ref{mfig6} shows the experimental (top), numerical (middle) and theoretical (bottom) instability windows for the new structure, in which the $1$ mm glass slides are sandwiched between $3.1$ mm air windows.  The longer spatial period in the structure results in a smaller spacing between the instability bands in Fourier space. Once again, we obtain good quantitative agreement between experiment, numerics, and theory with respect to the locations of the instability bands. In the experiments, the peaks of the first two bands are at $k=5.3\times 10^{-3}$ and $k=1.46\times 10^{-2}$, with values of $R=2.13$ and $R=1.35$, respectively; a third band appears near $k=2.05\times 10^{-2}$, with $R=1.03$. We note that because of the larger air gaps, the instability windows shift towards lower wavenumbers.
Observe, however, that there is a slight disparity in the location of the very weak (in the experiments) third band between the analytical, numerical, and experimental results. This may be attributable to the very weak growth rate of the instability in conjunction with our quasi-1D approximation versus the fully three-dimensional spatio-temporal nature of the experiment.

\begin{figure}[tbp]
\centerline{(a)\includegraphics[width=8.0cm]{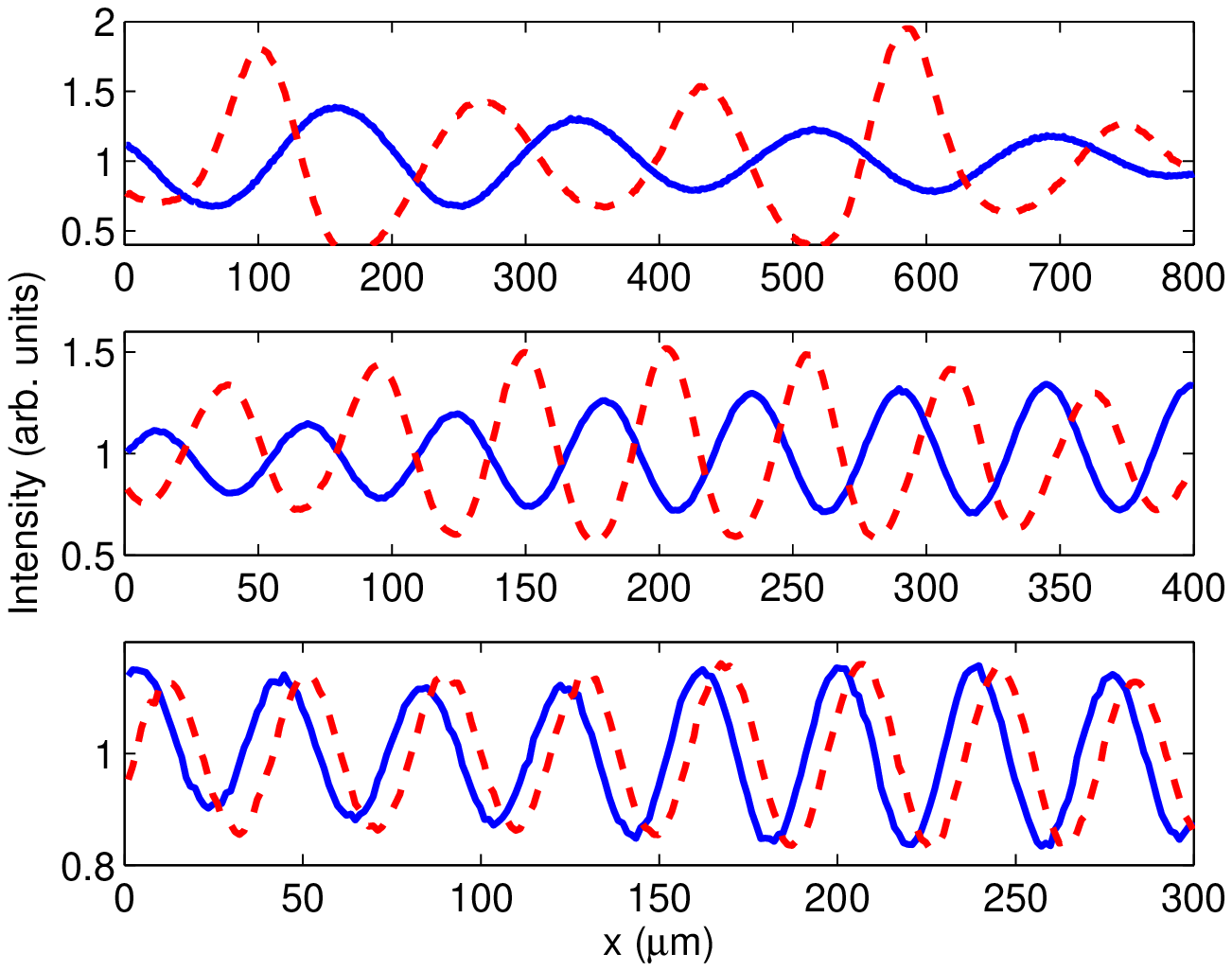}} \centerline{(b)\includegraphics[width=8.0cm]{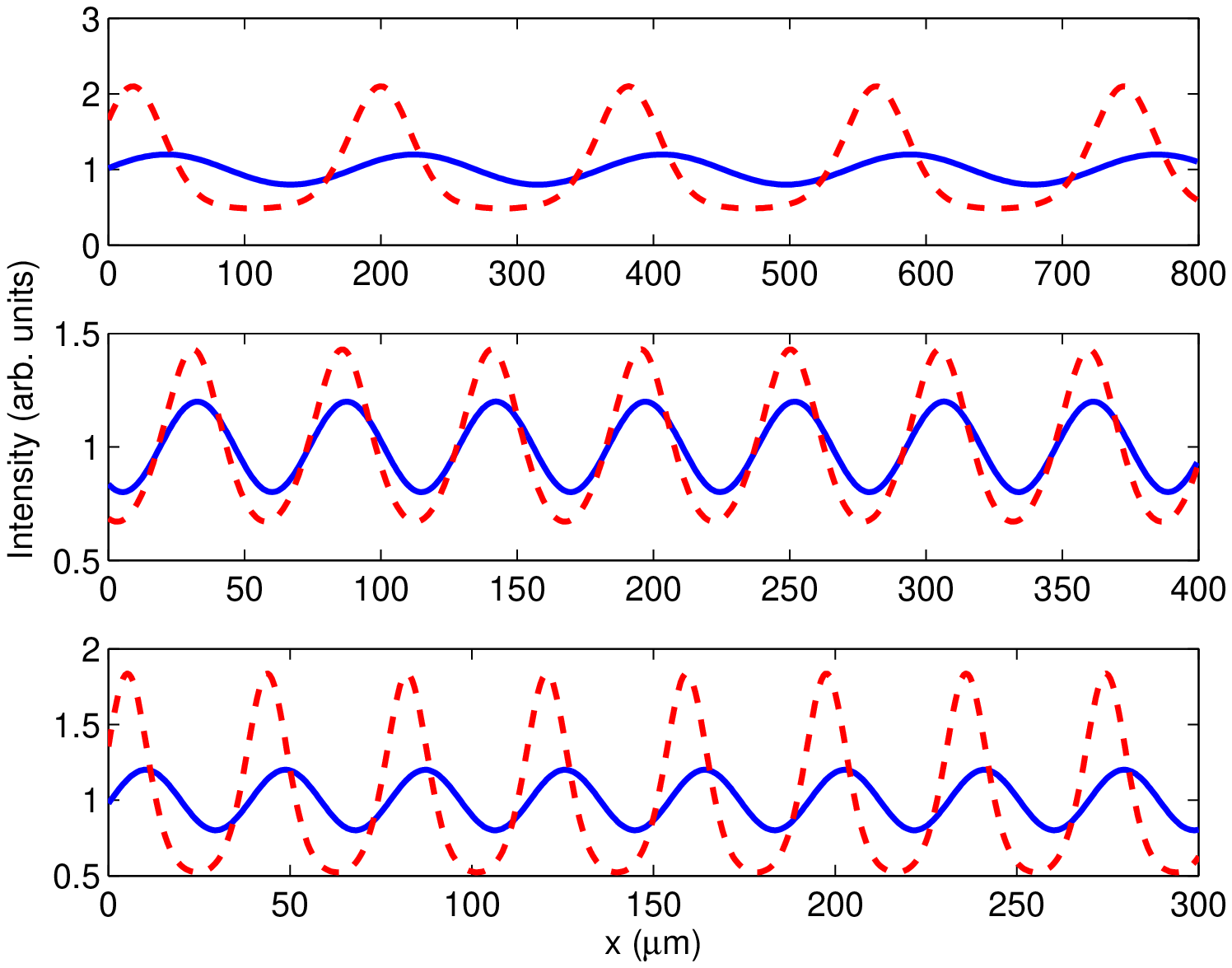}} \caption{(Color online) Same as Fig.~\ref{mfig3} but at the end of the propagation ($z = 21.5$ mm) for the 1 mm glass--3.1mm air configuration. The wavenumbers in the experimental plots are $k = 0.0042$ [top panels of (a) and (b)], $k = 0.0146$ (middle), and $k = 0.0205$ (bottom). The first two wavenumbers are the same in the numerical plots, but we use $k = 0.0208$, the peak of the third numerical band, for the last plot.  These wavenumbers occur, respectively, in the first instability band, the second instability band, and the third instability band. As before, the dashed curves are for high intensity $(I_{P2})$, and the solid ones are for low intensity $(I_{P0})$. } \label{inter4k}
\end{figure}

Figure \ref{inter4k} shows the normalized one-dimensional intensity pattern at the output of the NLM for $I_{P2}$ (dashed curve) and $I_{P0}$ (solid curve).  For both the experimental [Fig.~\ref{inter4k}(a)] and theoretical [Fig.~\ref{inter4k}(b)] results, the three panels are for $k=0.0044$ (top), which lies in the first instability band; $0.0146$ (middle), which lies in the second instability band; and $k=0.0205$ (bottom) [$k=0.0208$ for the numerical results], which lies in the third instability band. Once again, we clearly observe an increase in the modulation depth in the instability bands, whereas for the modulationally stable case, such an increase is absent. In fact, the experiment may even suggest a corresponding decrease, which does not appear in the numerical simulations. We believe this decrease results from the non-ideal conditions of the experiments---particularly because the modulation of the input beam is not purely sinusoidal and the beam itself is not a plane wave. The third band shows only a weak instability. Experimentally, the larger angle between the two beams (leading to the high wavenumbers of this band) reduces the interaction between them because of the limited temporal and spatial overlap, thereby decreasing the strength of the nonlinearity.

\begin{figure}[tbp]
\centerline{\includegraphics[width=10.0cm]{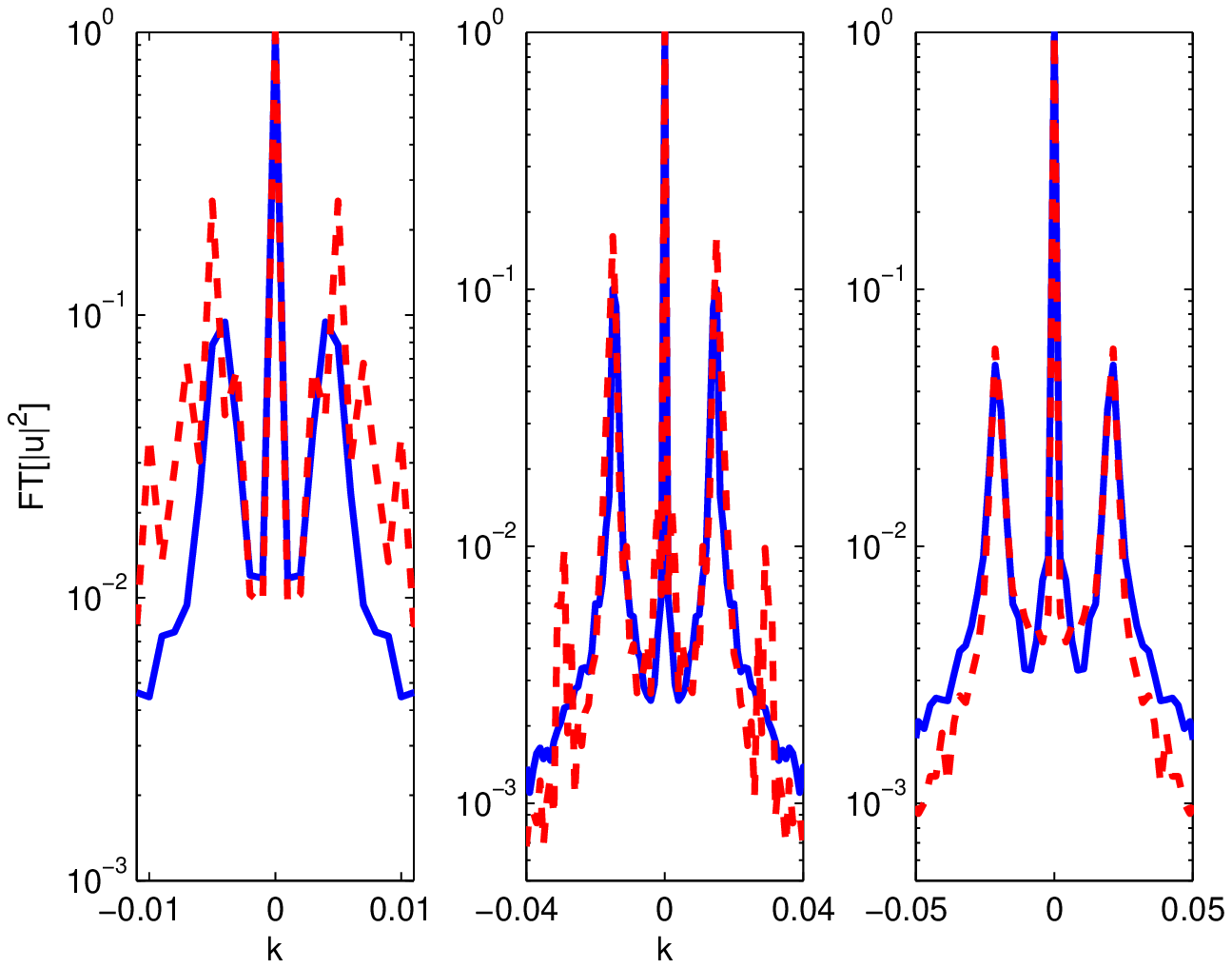}}
\centerline{
\includegraphics[width=10.0cm]{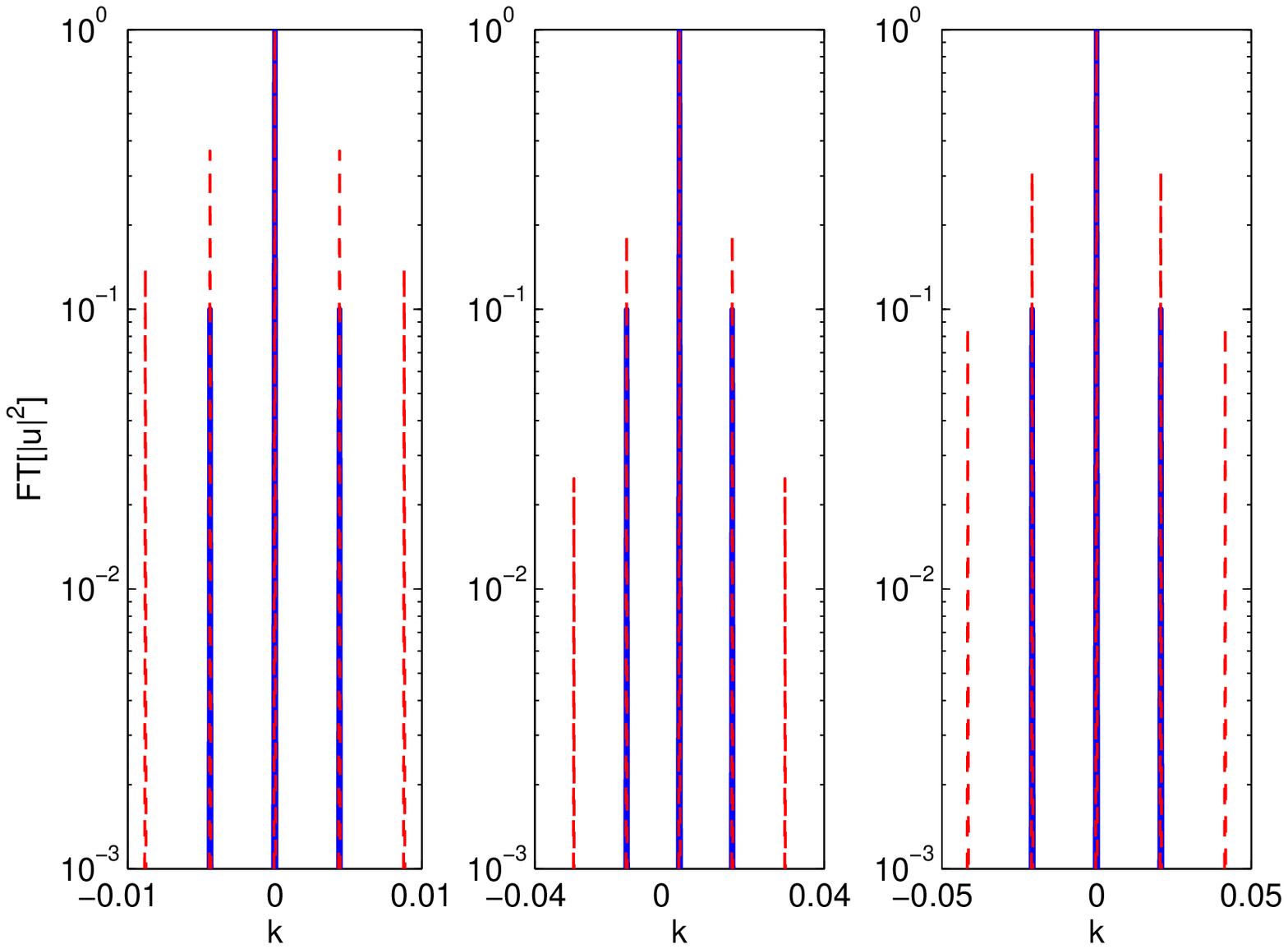}
} \caption{(Color online) Same as Fig.~\ref{mfig4} but at the end of the propagation ($z = 21.5$ mm) for the 1 mm glass--3.1mm air configuration.  As before, the experimental results are shown in (a) and the numerical ones are shown in (b).  The left panels show the results for wavenumber $k=0.0044$ (first instability band), the middles panels are for $k = 0.0146$ (second band), and the right panels are for the third band ($k=0.0205$ for the experiments and $k=0.0208$ for the numerical simulations). As before, the dashed curves are for high intensity $(I_{P2})$  and the solid ones are for low intensity $(I_{P0})$.} \label{fourier4k}
\end{figure}

In Fig.~\ref{fourier4k}, we show the Fourier transforms of the experimental (top panels) and numerical (bottom panels) intensity patterns at the output of the layered Kerr medium with the $3.1$ mm air gaps. The appearance of higher spatial harmonics of the initial modulation in the regions of instability is again evident both experimentally and numerically, especially in the first two modulationally unstable bands. For the third band, the phenomenon is very weak, as discussed above. From left to right, the panels depict the results for wavenumbers in the first ($k = 0.0044$), second ($k = 0.0146$), and third ($k = 0.0205$ in the experiments and $k=0.0208$ in the numerical simulations) unstable bands.


\subsection{Discussion}
In this section, we discuss the evolution of $u(x,z)$ and elaborate on several of the points mentioned in previous sections. In particular, we discuss the evolution of the diagnostic $R=R(z)$, previously reported only at the output of the layered medium (in order to compare with experiments). We also discuss how the different intensities (and hence different effective nonlinearities) influence our experimental results, in terms of the diagnostic $R$ as a function of wavenumber. For that same dependence ($R$ as a function of wavenumber $k$), we consider the ripples previously discussed for the stable region and how their relative amplitudes compared to the peak heights in the instability regions vanish as the propagation distance (that is, the number of propagation periods) increases.  Finally, we validate the assumption (used in our theoretical analysis) of substituting the weak variation in the dispersion by its average by comparing the direct evolution results between the true dynamics and the average-dispersion ones.

\begin{figure}[tbp]
\centerline{\includegraphics[width=4.7cm]{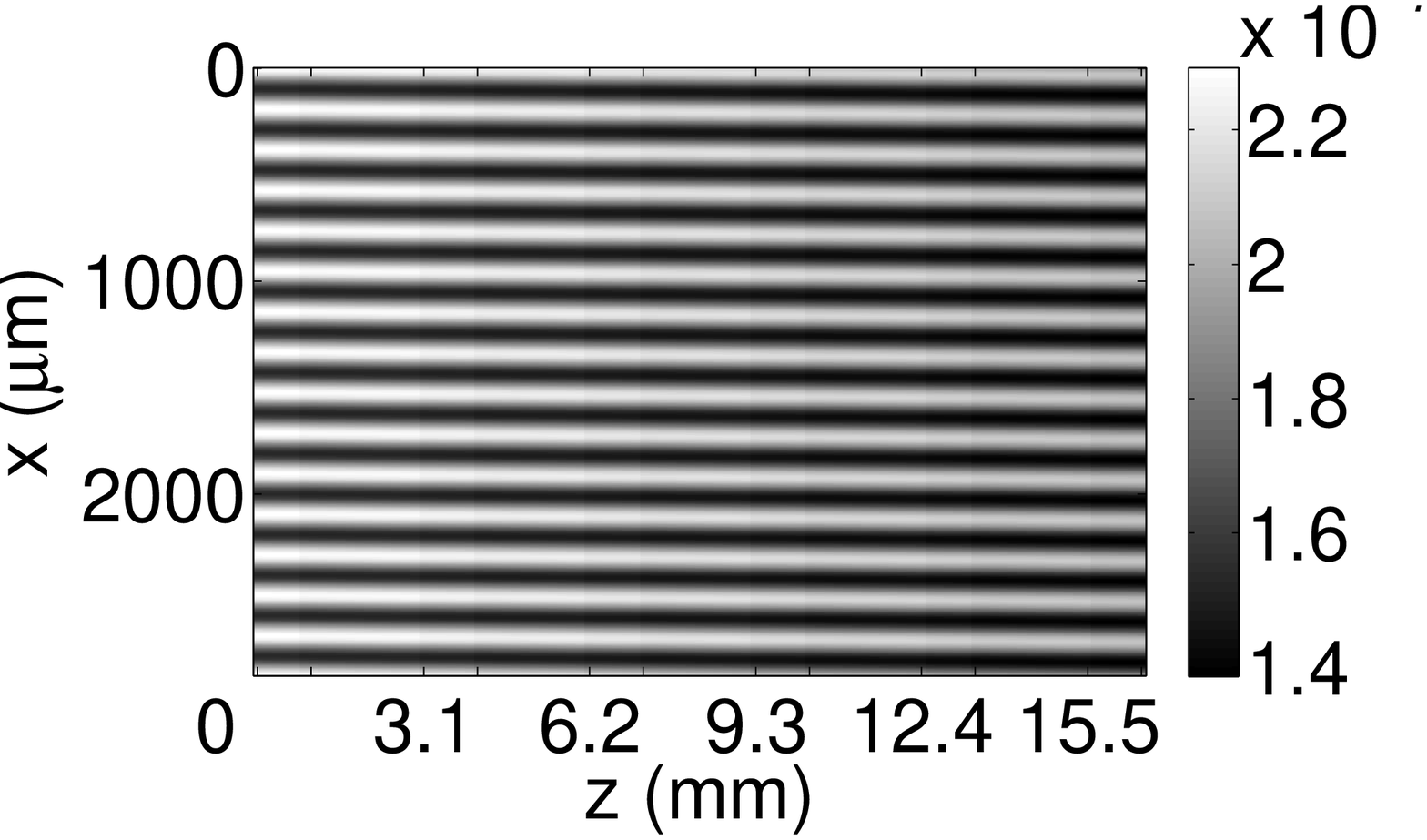}
\includegraphics[width=4.7cm]{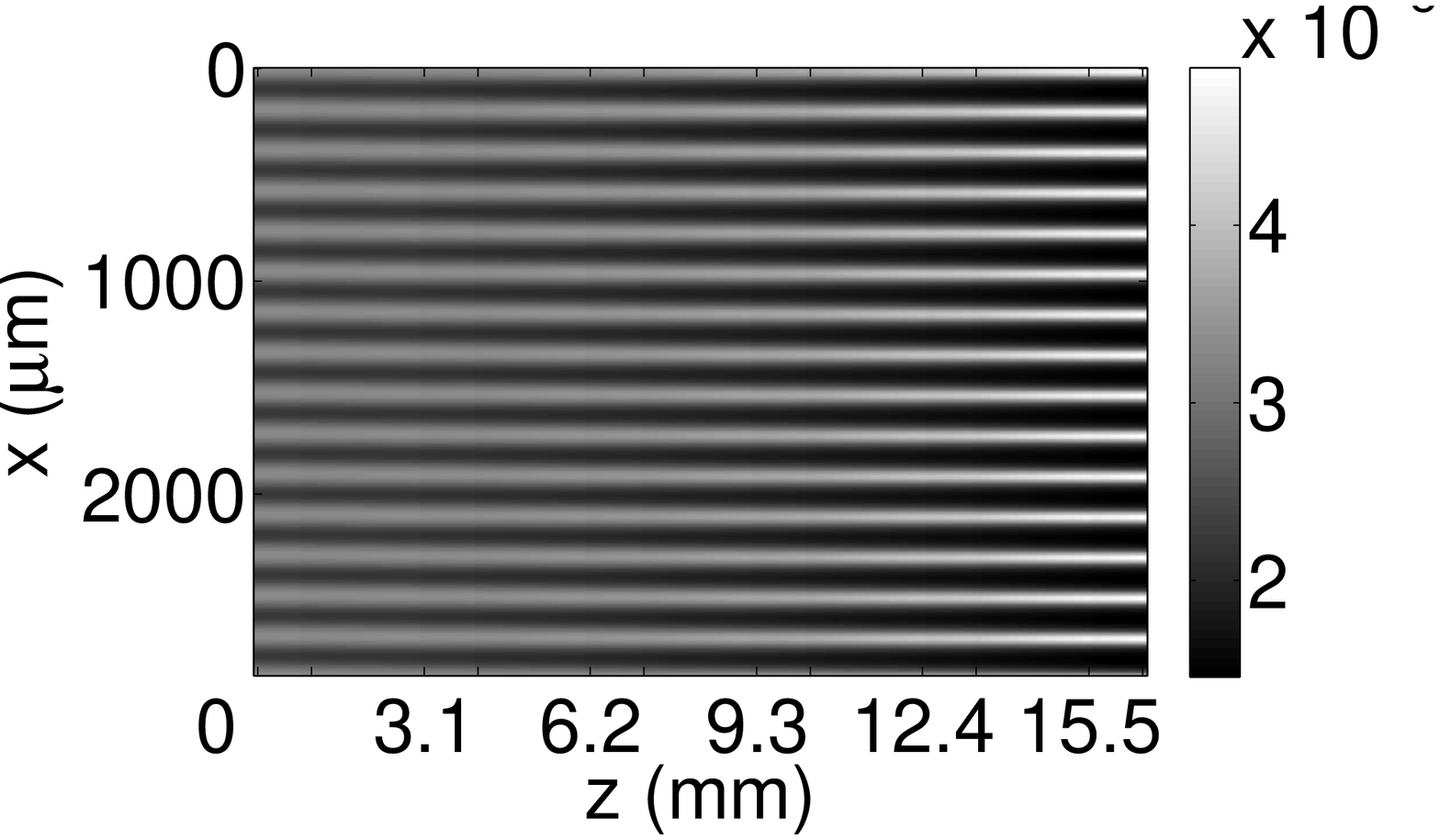}}
\centerline{\includegraphics[width=4.7cm]{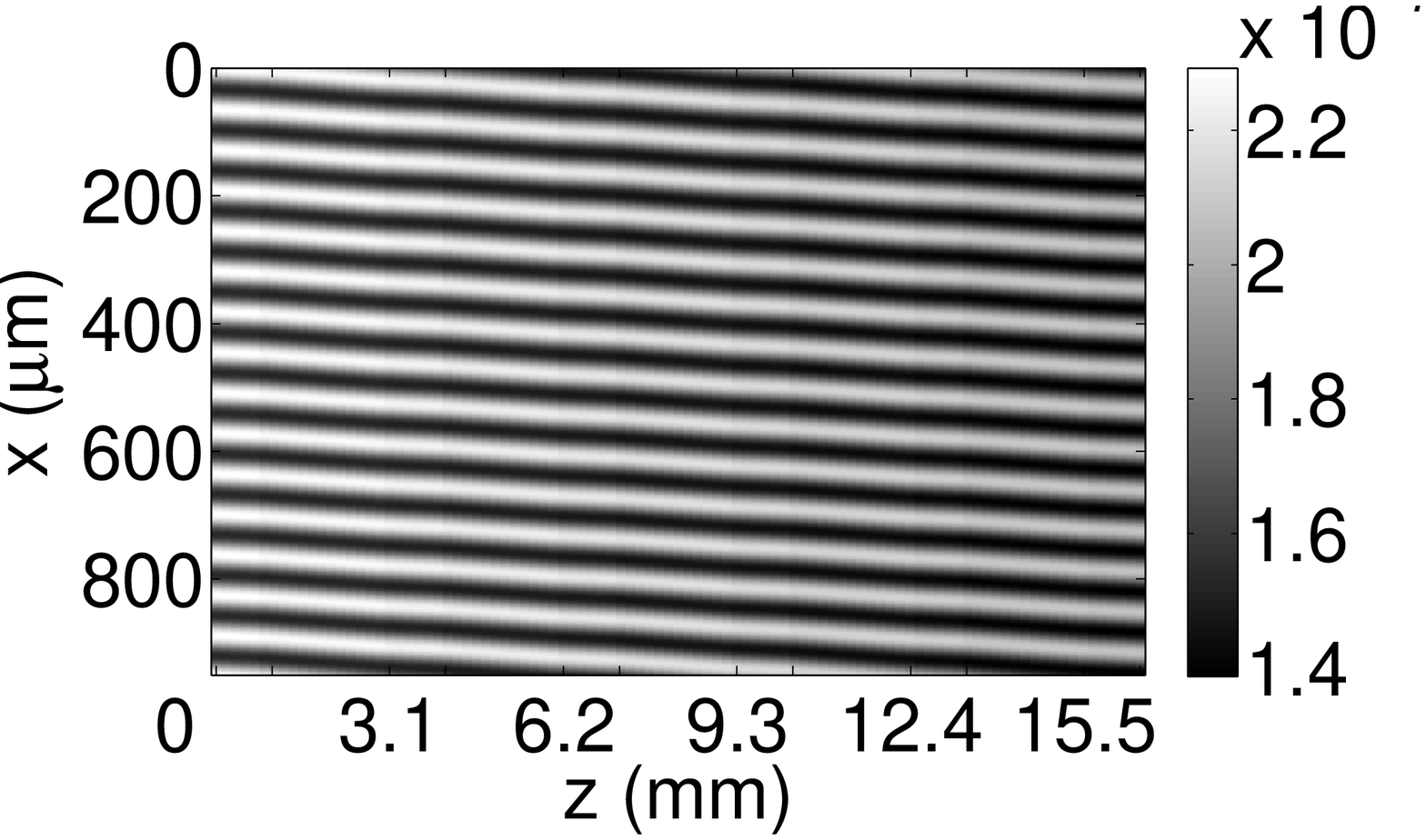}
\includegraphics[width=4.7cm]{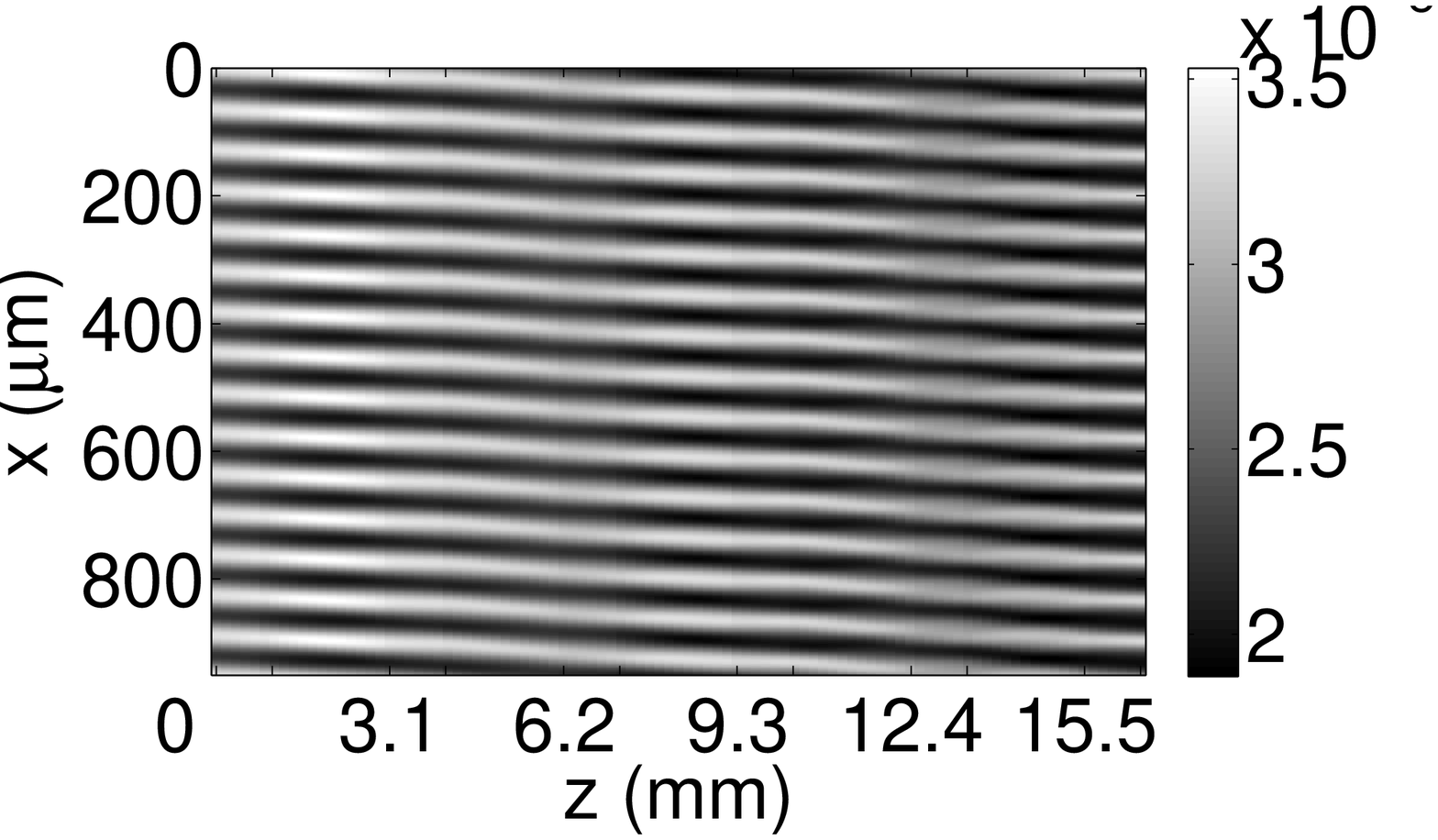}}
\centerline{\includegraphics[width=4.7cm]{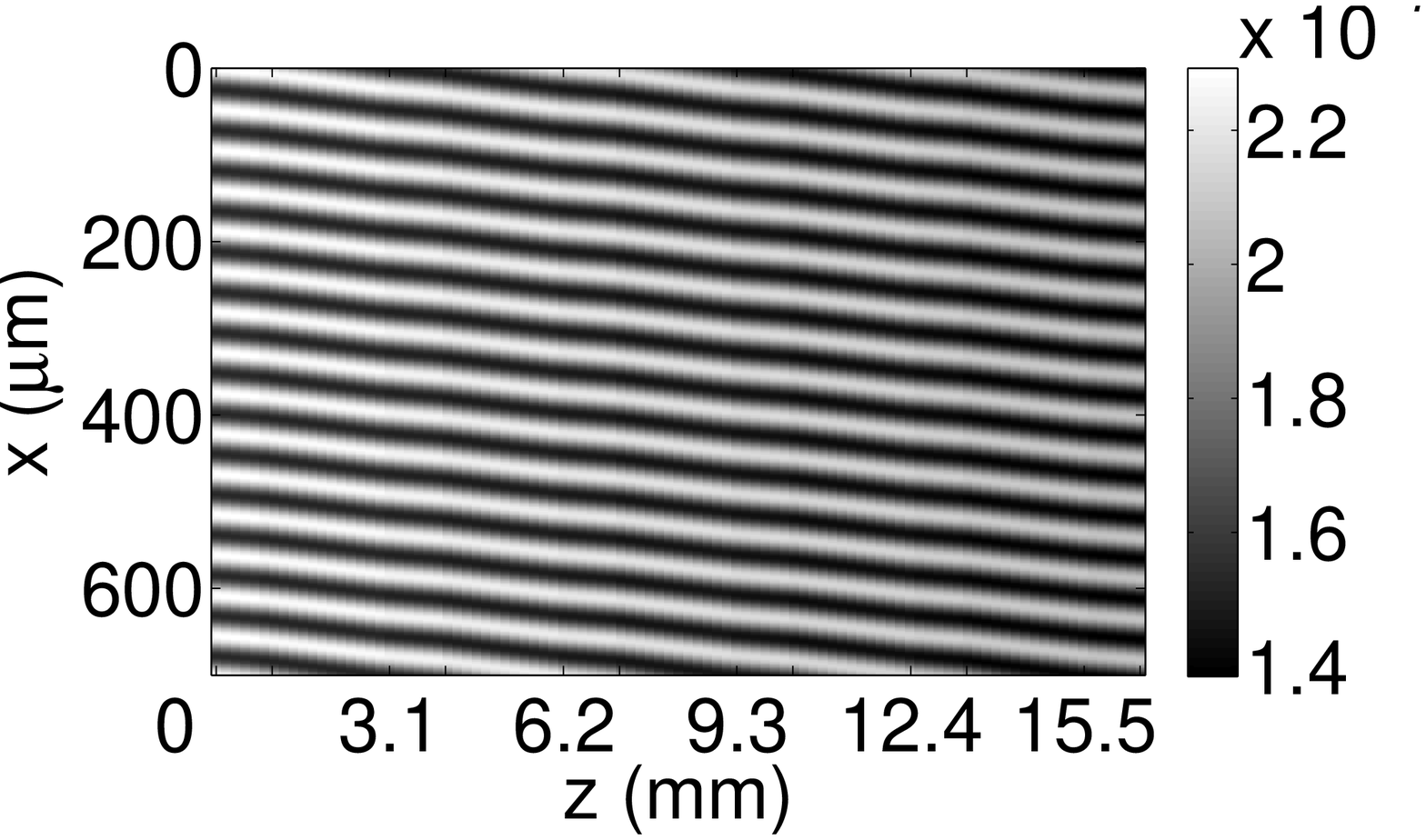}
\includegraphics[width=4.7cm]{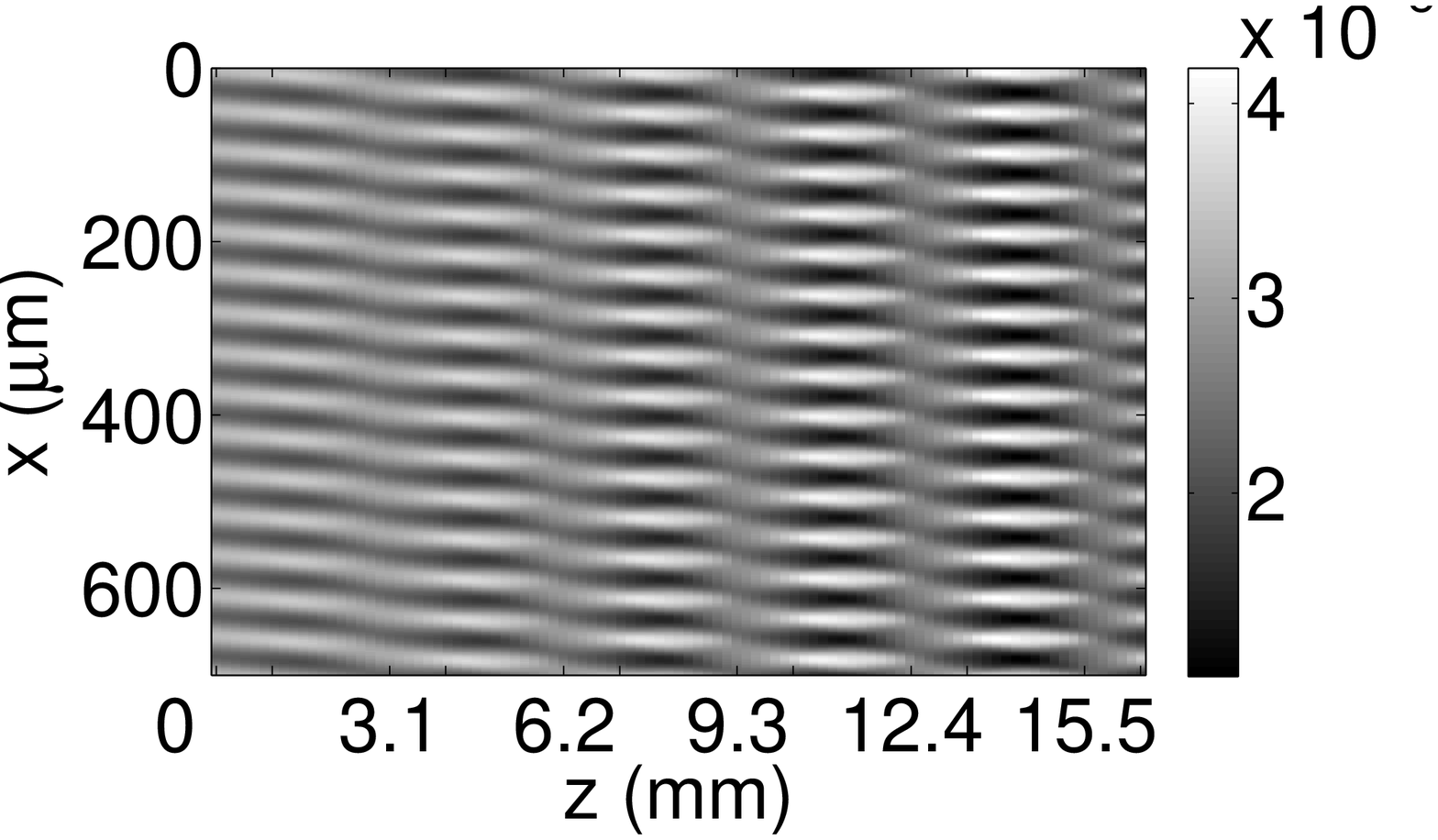}}
\caption{Evolution of the intensity $|u(x,z)|^2$ (in arbitrary units) in layered Kerr media consisting of 1 mm glass--2.1 mm air layers. We simulate initial wavefunctions of both low intensity (left) and high intensity (right) plane waves with a small modulation of wavenumber $k = 0.0042$ (top), $k=0.0126$ (middle), and $k=0.0170$ (bottom). For each of the numerical simulations, we consider domains with 15 periods.  The ticks on the horizontal axes indicate the glass--air interfaces, and the labeled values indicate the left edges of the glass slides.} \label{evo42}
\end{figure}

In Fig.~\ref{evo42}, we show contour plots of the intensity $|u(x,z)|^2$ for our numerical simulations (for the layered medium with 1 mm glass slides sandwiching 2.1 mm air gaps) of low-intensity and high-intensity initial wavefunctions at $k = 0.0042$ (first instability band), $k = 0.0126$ (stable region), and $k = 0.0170$ (second instability band). For the stable region, the nonlinearity has only a small effect on the evolution of the interference pattern. For propagation in the instability bands, we observe an increase in modulation depth in both cases but with marked differences in the evolution.  For the lower wavenumber, the modulation grows continuously and the orientation of the pattern remains fixed.  The most interesting behavior is observed in the second instability band, where the evolution of the pattern is markedly different. The modulation depth increases and decreases periodically, with a net increase after each period. The orientation of the pattern shifts dramatically between the different materials, possibly suggesting a periodic change in the relative angle between the two beams.

\begin{figure}[tbp]
\centerline{\includegraphics[width=10.0cm]{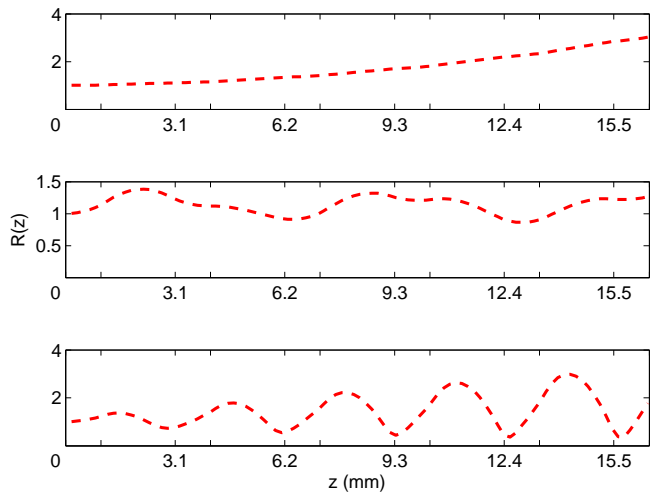}
}
 \caption{(Color online) Evolution of the MI diagnostic $R = R(z)$ as a function of the propagation
distance (in $\mu$m) for the configuration with 1 mm glass slides sandwiching 2.1 mm of air. The wavenumbers are $k = 0.0042$ (top), $k = 0.0126$ (middle), and $k = 0.0170$ (bottom). The ticks on the horizontal axis indicate the glass--air interfaces, and the values labeled on the axis indicate the left edges of the glass slides.}
\label{diagnostic}
\end{figure}

We now discuss in greater detail the dynamical dependence of the diagnostic $R$ on the propagation distance $z$ (see Fig.~\ref{diagnostic}).
For the case of 2.1 mm air gaps, we show examples in the first instability band ($k = 0.0042$), the second instability band ($k = 0.0170$), and in the region between the two instability bands ($k = 0.0126$).  For the second unstable region, observe that $R(z)$ oscillates but with an increasing amplitude.
Note additionally that the peak of the oscillation is not at the boundary between the two layers but rather near the half-period point.
We observe similar features for the configuration with 3.1 mm air gaps.

\begin{figure}[tbp]
\centerline{\includegraphics[width=10.0cm]{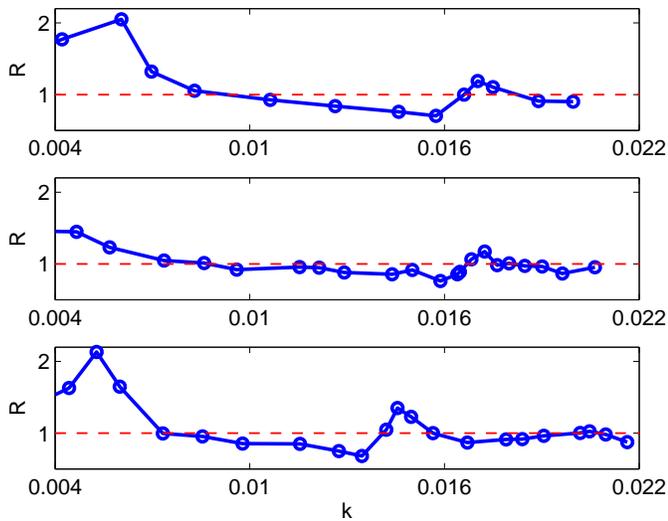} }
 \caption{(Color online) Experimental measurement of $R$ versus the dimensionless wavenumber $k$.
The top panel corresponds to the 1 mm glass--2.1 mm air configuration with intensity $I_{P2}$, the middle panel corresponds to the same structure with lower intensity $I_{P1}$, and the bottom panel corresponds to the structure with 1 mm glass--3.1 mm air and intensity $I_{P2}$. Observe in this last panel the leftward shift of the MI bands and the presence of a third band.} \label{mfig5}
\end{figure}

Figure \ref{mfig5} shows the experimentally-measured instability windows [$R = R(k)$] for the 2.1 mm air--1 mm glass configuration with an initial beam of high intensity $I_{P2}$ (top), one with lower intensity $I_{P1}$ (middle), and for the 3.1 mm air--1 mm glass configuration with high intensity (bottom). For the lower intensity case, the peaks are smaller because of the weaker nonlinearity, in consonance with the theoretical prediction.  For the structure with larger air gaps, the instability bands shift towards lower $k$.  Moreover,
as seen by Eq.~(\ref{nls}),
one can obtain more instability windows by increasing the widths of the air gaps further.  However, this is very difficult to achieve in experiments. The interaction length of the two beams is limited by their spatial and temporal overlap. In order to study structures with larger gaps, the beam diameter and pulse duration must be increased while keeping the same value of the intensity, which in our case was not possible due to the limited pulse energy.

\begin{figure}[tbp]
\centerline{\includegraphics[width=9.0cm]{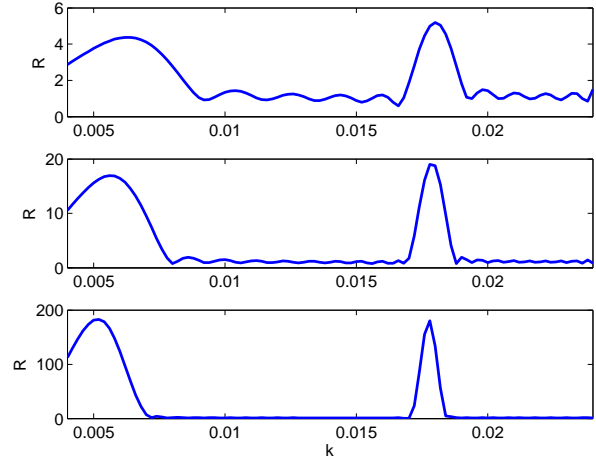} }
 \caption{(Color online) Numerical calculation of the instability bands (for the structure with 2.1 mm air gaps) for propagation with 6, 11, and 21 layers of glass (top, middle, and bottom panels, respectively). The heights of the ripples in the stable regions remain constant, whereas the heights of the peaks due to MI grow exponentially with propagation distance.
 } \label{mfig7}
\end{figure}

We now examine the diagnostic $R(k)$ for our direct numerical simulations as a function of propagation distance (number of propagation periods). Our aim is to explain the ripples in the stable regions in the middle panels of Figs.~\ref{mfig2} and \ref{mfig6}. We stress that these ripples do {\it not} correspond to an actual instability. Instead, they result from the fact that the simulations propagate over a finite number of periods.  We observed earlier that no such ripples appear in the instability windows computed using Bloch theory, which implicitly assumes that the number of periods in the propagation direction is infinite.  We thus expect the prominence of such ripples to decrease for direct numerical simulations with more layers of glass and air. As shown by the numerical instability peaks in Fig.~\ref{mfig7} for propagation distances with 6 (top panel), 11 (middle), and 21 (bottom) layers of glass (and 5, 10, and 20 sandwiched air gaps), this is indeed the case.  In these numerical experiments, we decreased the amplitude of the input reference wave ($\epsilon_0$) from $0.1$ to $10^{-3}$ to prevent a saturation in the growth of the modulation for the longer propagation distances. The simulations show that the amplitudes of the ripples remain essentially unchanged, whereas the MI peaks grow exponentially with distance, as expected from the theory. The position and periodicity of the ripples changes with the different propagation distances, which can be explained by the manner in which the growth of the reference beam evolves with propagation distance.


\begin{figure}[tbp]
\centerline{\includegraphics[width=10.0cm]{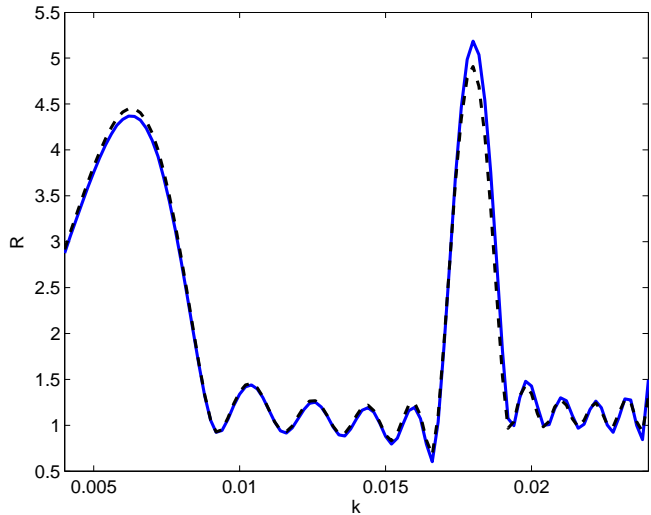} } \caption{(Color online) Instability windows for 1 mm glass--2.1 mm air configuration for piecewise constant $D(\zeta)$ (solid curve) and $D(\zeta) = \bar{D} = {\rm constant}$ (dashed curve).} \label{uniform}
\end{figure}

Finally, we revisit (as promised earlier) the assumption in our mathematical analysis of a uniform, mean-valued dispersion in our study of MI. To do this, we performed direct numerical simulations in which the coefficient of the Laplacian was uniform (assuming the values of the weighted averages used in the theory) rather than piecewise constant.  For the 1 mm glass--2.1 mm air configuration, the mean of $D(\zeta)$ is $\bar{D} = (1.5 + 2.1)/3.1 \approx 1.16$. For the 1 mm glass--3.1 mm air configuration, it is $\bar{D} = (1.5 + 3.1)/4.1 \approx 1.12$. As shown for the former configuration in Fig.~\ref{uniform} (we obtained similar results for the latter one), such changes result in almost no differences in the location of the instability windows and only small differences in the sizes of the instability peaks. We therefore assert that this provides an excellent approximation for determining the locations of the modulationally unstable wavenumber windows. The controllability and validity of our approximations can therefore be used to explain the very good quantitative agreement that we observe between our analytical, numerical, and experimental results, especially in light of the fact that there are no free/adjustable parameters in our model.




\section{Conclusions}

In the present work, we provided an experimental realization of the modulational instability (MI) in a medium that is periodic in the evolution variable. We also described the location of the instability bands {\it quantitatively} by investigating the Hill equation obtained from a linear stability analysis of plane-wave solutions of a nonlinear Schr\"odinger equation with piecewise constant nonlinearity coefficients.
In this case, the Hill equation becomes the Kronig-Penney model and thereby provides a direct association of the MI bands with the forbidden energy zones of that model. One of the unique features of the periodic medium in this respect (which can also be seen in the dispersion-managed nonlinear Schr\"odinger equation \cite{kutzmod,kumar}) is the opening of additional MI bands, such as the second and third bands discussed in detail in the present work. The precise location of the bands is chiefly determined by the details of the periodicity in the evolution variable (the thickness of the glass slides and the width of the air gaps).  We note in passing that our theoretical result can be further analyzed in the limit in which the glass slides are much wider than the air gaps (or vice versa), in which case one can infer, for example, that for the higher zones (indexed by $n$), the zone width is inversely proportional to $n$ \cite{krivchenkov}.  However, we did not pursue this case in detail, as it was not experimentally tractable. We compared our mathematical analysis for the modulationally unstable bands to both numerical and experimental findings (using Fourier-space diagnostics to elucidate the instability of the latter) and found very good agreement.  We also clearly observed higher spatial harmonics for modulationally unstable beams, revealing another characteristic trait of MI.

Additionally, the efficacy of layered Kerr media (and other instances of nonlinearity management) for examining interesting and important nonlinear dynamics goes far beyond the present work on MI. In a recent paper \cite{centurion}, we used a similar setup (alternating layers of glass and air) to stabilize an optical pulse using nonlinearity management, showing that it can potentially provide a lossless self-guiding mechanism.  Because both air and glass are focusing media, collapse or dispersion cannot be entirely prevented. In fact, in our setting, the presence of very weak dissipation always appears to favor the scenario of eventual dispersion of the pulse. Nevertheless, this occurs for propagation distances that are an order of magnitude larger than the typical ones of uniform media (and, moreover, the setting can be improved considerably).
We also captured our experimental results qualitatively (and, when appropriate, also quantitatively) by a $(2 + 1)$-dimensional nonlinear Schr\"odinger equation with a piecewise constant nonlinearity coefficient (with losses incorporated at the glass--air boundaries).
The very good agreement between theoretical, numerical, and experimental results both here and in the aforementioned previous work \cite{centurion} suggests a variety of extensions not only in the present setting of layered Kerr media but also in nonlinearity-managed Bose-Einstein condensates (whose mean-field dynamics are also governed by nonlinear Schr\"odinger equations). In particular, it will be very interesting to realize nonlinearity management in three-dimensional Bose-Einstein condensates and examine its impact on the stability of coherent structures---including both solitary waves and more complex entities such as vortices bearing topological charge (see, e.g., \cite{monte}).

These recent developments (especially in optics but also Bose-Einstein condensation) underscore the interest in further developing and expanding the mathematical theory pertinent to such settings.  In particular, it would be of interest to examine the well-posedness of such temporally-modulated settings (either at the level of the full model or at the level of its averaged variants), an avenue of research that is only starting to develop \cite{konotop1,atanas1}.

{\it Acknowledgements.} We acknowledge support from the DARPA Center for Optofluidic Integration (DP), the Caltech Information Science and Technology initiative (MC, MAP), the Alexander von Humboldt-Foundation (MC) and NSF-DMS-0204585, NSF-DMS-0505663, NSF-DMS-0619492, and NSF-CAREER (PGK).


\end{document}